\documentclass[trackchanges]{aastex701}

\usepackage{multirow}
\usepackage{array}
\usepackage{booktabs}
\usepackage{threeparttable}

\begin{document}

\title{HAWC J0630+186 could not be powered by PSR J0630+19}

\correspondingauthor{Xiaohong Cui, Bojun Wang}
\email{xhcui@bao.ac.cn, wangbj@bao.ac.cn}

\author[orcid=0000-0002-9434-4773]{Bojun Wang}
\affiliation{National Astronomical Observatories, Chinese Academy of Sciences, Beijing, China}
\email{wangbj@bao.ac.cn}

\author[orcid=0000-0002-6322-7582]{Xiaohong Cui}
\affiliation{National Astronomical Observatories, Chinese Academy of Sciences, Beijing, China}
\affiliation{Key Laboratory of Radio Astronomy and Technology, National Astronomical Observatories, CAS, Beijing, China}
\email{xhcui@bao.ac.cn}

\author[orcid=0000-0002-9815-5873]{Jiguang Lu}
\affiliation{National Astronomical Observatories, Chinese Academy of Sciences, Beijing, China}
\email{lujig@nao.cas.cn}

\author[orcid=0000-0002-5031-8098]{Heng Xu}
\affiliation{National Astronomical Observatories, Chinese Academy of Sciences, Beijing, China}
\affiliation{Key Laboratory of Radio Astronomy and Technology, National Astronomical Observatories, CAS, Beijing, China}
\email{hengxu@bao.ac.cn}

\author[orcid=0000-0002-9042-3044]{Renxin Xu}
\affiliation{Kavli Institute for Astronomy and Astrophysics, Peking University, Beijing, China}
\affiliation{Department of Astronomy, School of Physics, Peking University, Beijing, China.}
\email{r.x.xu@pku.edu.cn}

\begin{abstract}

3HWC J0630+186 is one of the very-high-energy gamma-ray sources listed in the third High-Altitude Water Cherenkov (HAWC) catalog; however, its origin remains unidentified. The only plausible counterpart is PSR J0630+19, which is offset from the center of 3HWC J0630+186. Several TeV halos around pulsars are currently considered the most dominant TeV-PeV \textbf{gamma-ray sources. PSR J0630+19 was first discovered in an} Arecibo survey as a normal-period pulsar, but its age and spin-down luminosity are not available, making it difficult to assess whether it is associated with 3HWC J0630+186. Using awarded observing time on the Five-hundred-meter Aperture Spherical radio Telescope (FAST), we conducted follow-up timing observations of PSR J0630+19 over more than one year. Through pulsar timing analysis, we obtained a more precise position and derived key parameters. These parameters indicate that PSR J0630+19 is an old pulsar with a spin-down luminosity too low to power the very-high-energy emission from 3HWC J0630+186.

\end{abstract}

\keywords{telescopes –pulsars: individual: PSR J0630+19 –ISM: individual objects: 3HWC J0630+186.
}

\section{Introduction}\label{sec:intro}

Discovering the origin of cosmic rays and elucidating their underlying acceleration mechanisms require precise, multi-wavelength observational data and analysis. The physical nature of numerous unidentified \textbf{galactic} sources detected in the very-high-energy (
$\geq$100GeV) gamma-ray ($\gamma$-ray) emissions band remains unclear. As one of the brightest gamma-ray sources, the Crab Nebula is powered by the ultra-relativistic electron-positron wind emanating from its central pulsar. It has been extensively observed from GeV to TeV energies \citep{2004ApJ...614..897A,2015ARNPS..65..245F,2015JHEAp...5...30A,2019ApJ...881..134A,2019PhRvL.123e1101A,2020NatAs...4..167H,2020ApJ...897...33A,2020A&A...635A.158M}. The discovery of \text{gamma-ray emissions with energies above PeV} \citep{2021Sci...373..425L,2021Natur.594...33C} \textbf{from this source implies that it is a PeV particle accelerator.}

The existence of TeV halos around pulsars is now firmly established, defining a new class of gamma-ray
sources with properties distinct from those of supernova remnants and pulsar wind nebulae (PWNe). Their emission is interpreted as originating from electrons and positrons freely diffusing in the surrounding interstellar medium (e.g. \citet{2020A&A...636A.113G, 2022FrASS...922100F, 2024icrc.confE.833L}). Extended TeV gamma-ray emission around the nearby Geminga pulsar was first revealed by the Milagro experiment \citep{2009ApJ...700L.127A} and later confirmed by the High Altitude Water Cherenkov (HAWC) observatory \citep{2017Sci...358..911A,2017ApJ...843...40A}.
Furthermore, \citet{2017PhRvD..96j3016L} \textbf{found a large fraction of 2HWC catalog sources coincident with} pulsars whose radio beams are not oriented towards Earth.

3HWC J0630+186 is one of the new TeV sources reported in \textbf{the} 3HWC catalog  \citep{2020ApJ...905...76A}, which notes that a significant fraction of its sources are candidate counterparts to pulsars listed in the ATNF catalog\footnote{https://www.at nf.csiro.au/research/pulsar/psrcat/} \citep{Manchester2005}. This source, detected in \textbf{a point-like source search}, is one of several excesses clustering near the Geminga pulsar. It has thus been hypothesized that 3HWC J0630+186 \textbf{and approximately four other nearby point-like sources} detected by HAWC collectively represent components of an extended halo surrounding the Geminga pulsar \citep{2017Sci...358..911A,2020ApJ...905...76A}. However, it remains unclear whether these discrete excesses correspond to a genuine, spatially structured halo surrounding the \textbf{Geminga} pulsar.

There is no GeV counterpart \textbf{reported in} 4FGL \citep{2020ApJS..247...33A} or known supernova remnant \textbf{in the} SNRCat \citep{2012AdSpR..49.1313F} to 3HWC J0630+186. The only possibly associated \textbf{known source is the} pulsar PSR J0630+19, located approximately $0^{^\circ}.95$ from the TeV source center. PSR J0630+19 was discovered by Arecibo 327 MHz drift pulsar survey \citep{Deneva2016} and exhibits a normal pulsar period. \textbf{Give the value of the period,}  key parameters such as its characteristic age and spin-down power \textbf{have not been determined} up to now. Consequently, it is unclear whether \textbf{PSR J0630+19} can supply the necessary energy to power the observed very-high-energy $gamma$-ray \textbf{emission observed by HAWC}.

\begin{deluxetable}{ccccccc}[htbp]
\tablecaption{Observation Log \label{tab:observation}}
\tablehead{
\colhead{No.} & 
\colhead{Date} & 
\colhead{MJD$^{a}$} & 
\multicolumn{2}{c}{Coordinate$^{b}$} & 
\colhead{Observation Mode} & 
\colhead{$T_{obs}^{c}$} \\
& 
\colhead{(yyyy-mm-dd)} & 
\colhead{} & 
\colhead{R.A.} & 
\colhead{Dec.} & 
\colhead{} & 
\colhead{(s)}
}
\startdata
1  & 2024-09-04 & 60556.97  & 06:30:04.00 & +19:37:00.0 & Tracking    & 3600 \\
2  & 2024-12-14 & 60657.75  & 06:30:04.00 & +19:37:00.0 & Tracking    & 7020 \\
3  & 2025-02-06 & 60712.62  & 06:30:04.00 & +19:37:00.0 & SnapShotDec & 2460 \\
4  & 2025-04-20 & 60785.35  & 06:30:05.33 & +19:35:03.8 & Tracking    & 1290 \\
5  & 2025-04-22 & 60787.31  & 06:30:05.33 & +19:35:03.8 & Tracking    & 1290 \\
6  & 2025-04-25 & 60790.36  & 06:30:05.33 & +19:35:03.8 & Tracking    & 1290 \\
7  & 2025-04-28 & 60793.32  & 06:30:05.33 & +19:35:03.8 & Tracking    & 1290 \\
8  & 2025-05-29 & 60824.25  & 06:30:05.33 & +19:35:03.8 & Tracking    & 1290 \\
9  & 2025-06-28 & 60854.20  & 06:30:05.33 & +19:35:03.8 & Tracking    & 1290 \\
10 & 2025-08-02 & 60889.06  & 06:30:05.33 & +19:35:03.8 & Tracking    & 1200 \\
11 & 2025-09-03 & 60920.95  & 06:30:05.33 & +19:35:03.8 & Tracking    & 1167 \\
12 & 2025-10-04 & 60951.90  & 06:30:05.33 & +19:35:03.8 & Tracking    & 1167 \\
\enddata
\tablenotetext{a}{At exposure start.}
\tablenotetext{b}{Coordinates of beam M01.}
\tablenotetext{c}{Observation time in seconds.}
\end{deluxetable}

To investigate the origin of the very-high-energy $\gamma$-ray emissions from 3HWC J0630+186, we used the Five-hundred-meter Aperture Spherical radio Telescope (FAST), which offers unprecedented sensitivity \citep{2006ScChG..49..129N,2019fast}, to conduct follow-up pulsar observations of the nearby source PSR J0630+19. Owing to its unparalleled sensitivity as the world's largest single-dish radio telescope, FAST has profoundly advanced pulsar discovery and timing. This is demonstrated by its key survey projects, including the Commensal Radio Astronomy
FAST Survey (CRAFTS; \citet{2018IMMag..19..112L}), the Galactic
Plane Pulsar Snapshot Survey (GPPS; \citet{2021RAA....21..107H}),
and the FAST Globular Cluster Pulsar Survey \citep{2021ApJ...915L..28P}. Precise follow-up timing \textbf{observations of pulsars} (e.g., \citet{2024MNRAS.528.6761L, 2025ApJ...991..231Y}) enable the measurements of their period derivatives. These measurements are crucial as they yield key physical parameters, including the characteristic age and spin-down energy loss rate of \textbf{pulsars}.

In this paper, we describe the FAST observations and data reduction for PSR J0630+19 in Section 2. Section 3 presents the results of the timing analysis and the derived pulsar parameters. We provide a comprehensive discussion of these results in Section 4, followed by a concluding summary in Section 5.

\section{Observation and Data analysis}\label{sec:data}

To confirm the pulsar signal reported by \citet{Deneva2016}, we observed PSR J0630+19 with FAST on September 4 and December 14, 2024, with integration times of 1.0 and 2.0 hours, respectively. We employed the 19-beam L-band receiver in tracking mode, covering a frequency range of 1.0–1.5 GHz with 4096 channels \citep{2019fast}. Data from all 19 beams were recorded with a temporal resolution of 49.152 $\mu$s and in full Stokes polarization. We conducted blind periodicity searches using the PulsaR Exploration and Search TOolkit (\texttt{PRESTO}) software package \citep{Ransom2001}. Radio frequency interference (RFI) mitigation was performed using routine \textbf{rfifind} provided in \texttt{PRESTO}. The dispersion measure (DM) search range was 0--500 pc cm$^{-3}$,
where the upper limit significantly exceeds the DM value reported in \citet{Deneva2016}. Data were de-dispersed using steps determined by \textbf{DDplan} routine in \texttt{PRESTO}: 0.2 pc cm${^-3}$ below 376 pc cm${^-3}$ and 0.3 pc cm${^-3}$ above. Additionally, Fourier-domain acceleration search was performed with the parameter $z_{max}=200$.

\begin{figure}[b]
    \centering
    \includegraphics[width=\textwidth]{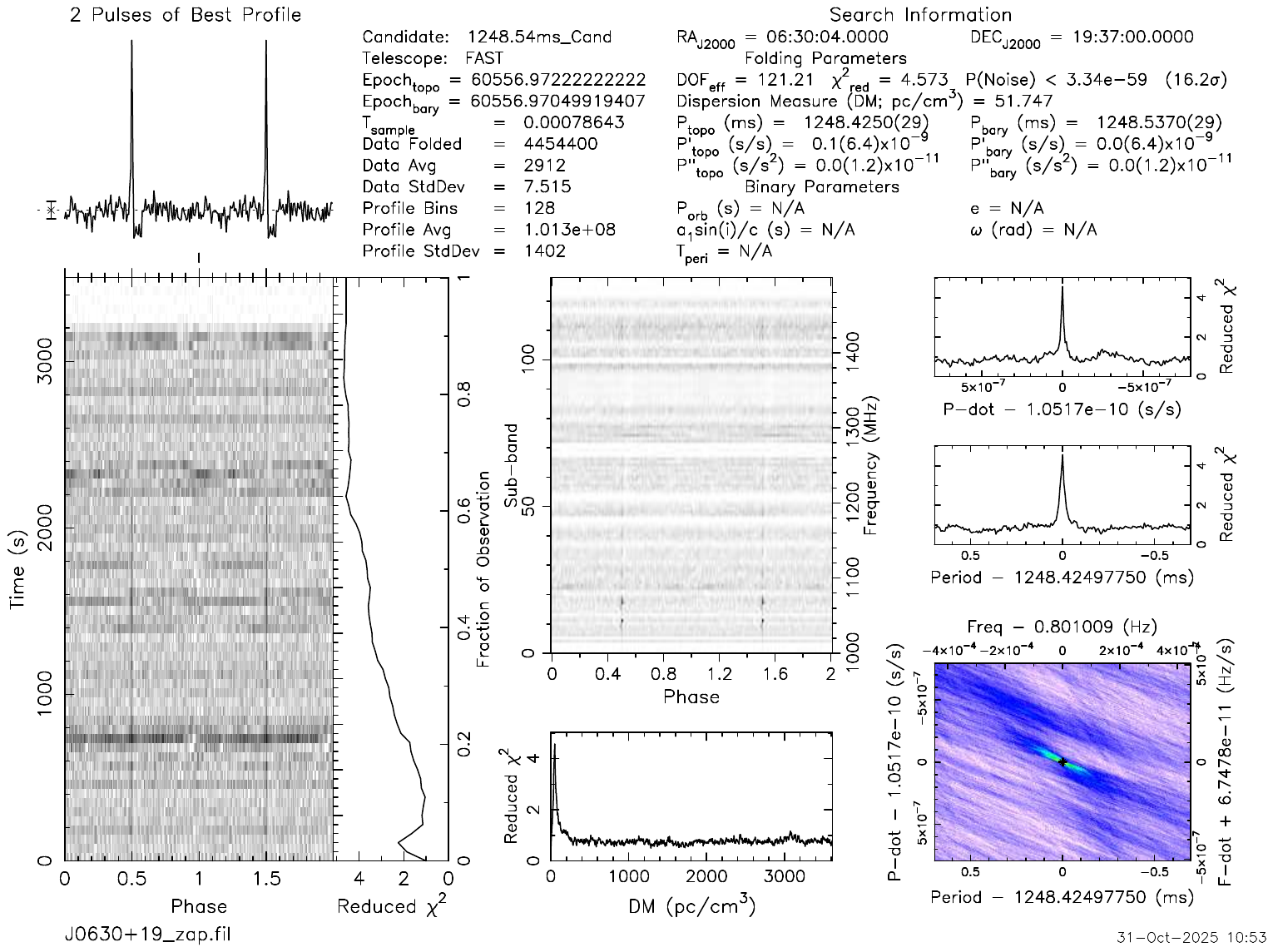}
    \caption{Radio pulses detected from PSR J0630+19 on September 4, 2024 is presented in the form of diagnostic plot generated using the \textbf{prepfold} routine in the \texttt{PRESTO} software package. The validity of derived parameters can be found in the top-right section of the plot, where the $\chi^{2}$ is a function of DM, $P$, and $\dot{P}$, respectively. The confidence contours for $P$ and $\dot{P}$ are displayed in the bottom-right section. In the central portion, the averaged pulse profile is shown as a function of observing frequency.}
    \label{fig:presto}
\end{figure}

As a point source, the position and its uncertainties of PSR J0630+19 were given in \citet{Deneva2016}. The positional uncertainties \textbf{were reported as} $7'.5$ in both Right Ascension (R.A.) and Declination (Dec.), larger than the angular resolution of FAST at 1.4 GHz ($\sim 2'.9$). After the radio pulses from J0630+19 were re-discovered and \textbf{confirmed, we 
used} the same receiver of FAST in $SnapShotDec ~mode$ to \textbf{obtain} 
a more accurate position for PSR J0630+19. This $SnapShotDec ~mode$ is designed to fill the gaps between the 19 beams by tracking observations of four pointing in a sequence. Each pointing in our snapshot observation lasted 10 minutes, and data from all beams were recorded with the same temporal resolution and polarization information as \textbf{in 
previous observations}.

In order to obtain high-precision timing solution for PSR J0630+19, we conducted multiple follow-up observations using the newly calculated coordinates, as listed in Table~\ref{tab:observation}. The updated coordinates allowed us to achieve a higher signal-to-noise ratios (SNR) with shorter integration time, thereby significantly conserving our observational time. For each observation epoch, raw data were folded using \textsc{DSPSR} software package \citep{vanStraten2011} into pulse profile with 2048 phase bins. Then we performed RFI \textbf{mitigation 
following} two steps: First, the two outermost 20-MHz bandwidths were removed (1000–1020 MHz and 1480–1500 MHz) due to sensitivity loss and phase instability. Second, the remaining spectral channels were cleaned using a median-filter algorithm. As a result, approximately 10\% to 30\% of the frequency channels were zapped for different epoch. Then we use routine \textbf{pac} from \texttt{PSRCHIVE} \citep{Hotan2004} software package to do the polarization calibration. We calculated the pulse time of arrivals (ToAs) by cross-matching each profile with a noise-free template \textbf{profile 
obtained} by fitting a set of Gaussian functions to the integrated profile formed from summing several of the highest-SNR observations. We \textbf{used} the Fourier domain Markov chain Monte Carlo ToA estimator to calculate \textbf{ToAs,  
implemented} in routine \textbf{$pat$} provide by \texttt{PSRCHIVE} software package. \textbf{In order to measure the DM in timing procedure, the data of each observation were divided into 4 frequency channels. We then measured the ToAs in each channel.} We use the \texttt{TEMPO2} \citep{Hobbs2006} software package to get the final timing solution and timing residuals.

\begin{figure}[htbp]
    \centering
    \includegraphics[width=0.9\textwidth]{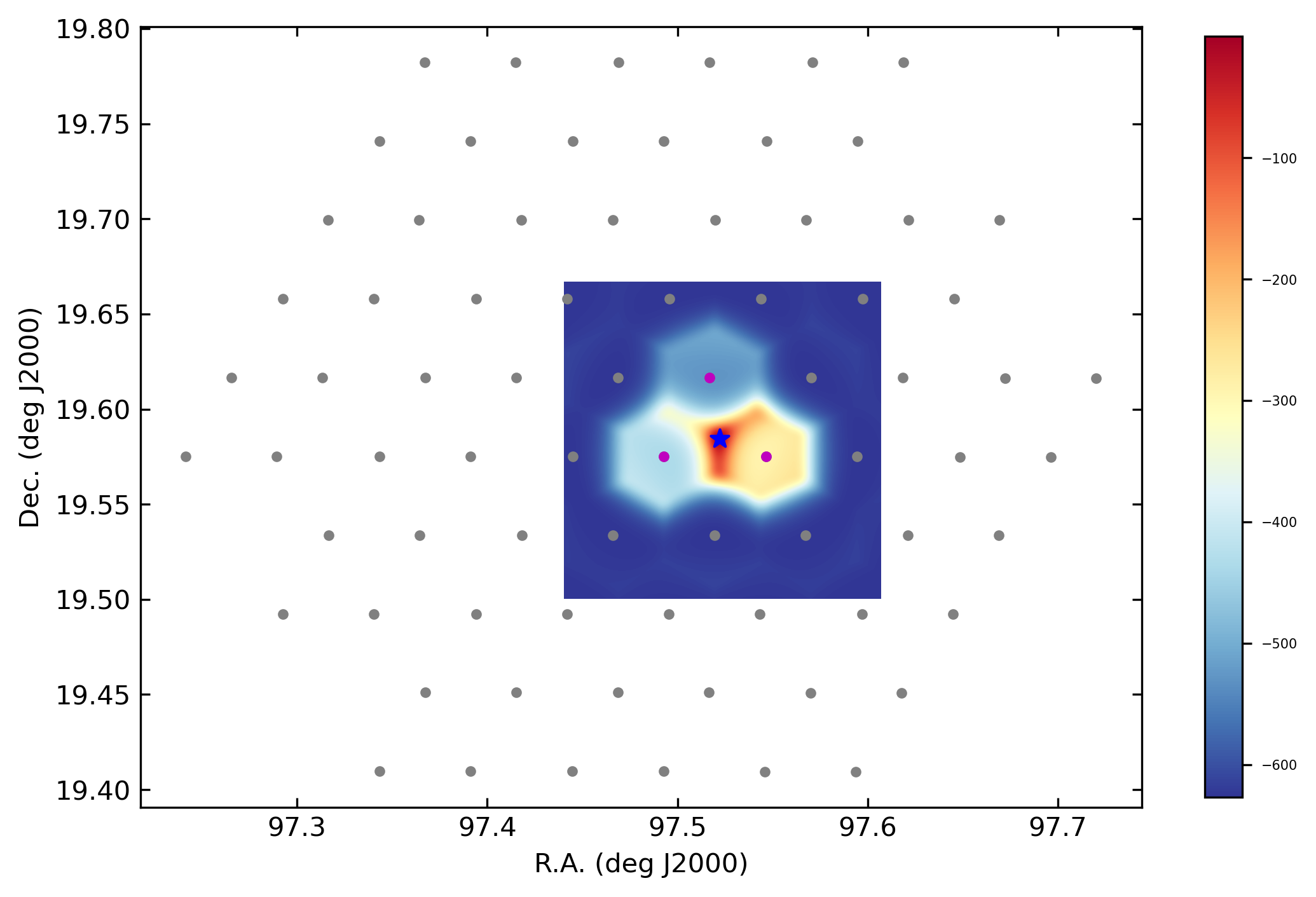}
    \caption{SnapShotDec observation of PSR J0630+19. The colormap represents the likelihood. All the dots (including gray and purple dots) are the beam positions. Three purple dots correspond to the three beams that detected pulse signals while the blue pentagram indicates the newly calculated coordinates for PSR J0630+19, as R.A.=06:30:05.33, Dec.=+19:35:03.8.}
    \label{fig:loc}
\end{figure}

\section{Results} \label{sec:result}

\subsection{Confirmation and Localization of PSR J0630+19}

During the two observation epochs on September 4 and December 14, 2024, we successfully detected radio pulses in beam M01 (as shown in Figure \ref{fig:presto}), with SNR of $\sim16.2\sigma$ and $\sim39.1\sigma$, respectively. The periods of the two signals were both \textrm{$\sim1248.54$ ms}, and the DM values were  $\sim51.7$ pc cm$^{-3}$, which are consistent with the values reported in \citet{Deneva2016}.

In our $SnapShotDec ~Mode$ observation on February 6, 2025, the pulse signals were detected both in the first pointing of M01, the forth pointing of M01 and the third pointing of M02. We folded the raw data from these three beams using the same integration time to generate pulse profiles. Assuming that the intensity of pulse profiles remained constant over a half-hour and based on the coordinates of three beam pointing that detected the pulse signals and the SNR of the detected pulses (see in Table~\ref{tab:loc}), we calculated a new \textbf{position} 
for PSR J0630+19, as R.A.=06:30:05.33, Dec.=+19:35:03.8 (as shown in \textbf{Fig.~\ref{fig:loc}). The} pointing accuracy of the feed receivers is $16\arcsec$ \citep{2019fast}.

\subsection{Timing and Properties of PSR J0630+19}

\begin{figure}
    \centering
    \includegraphics[width=\textwidth]{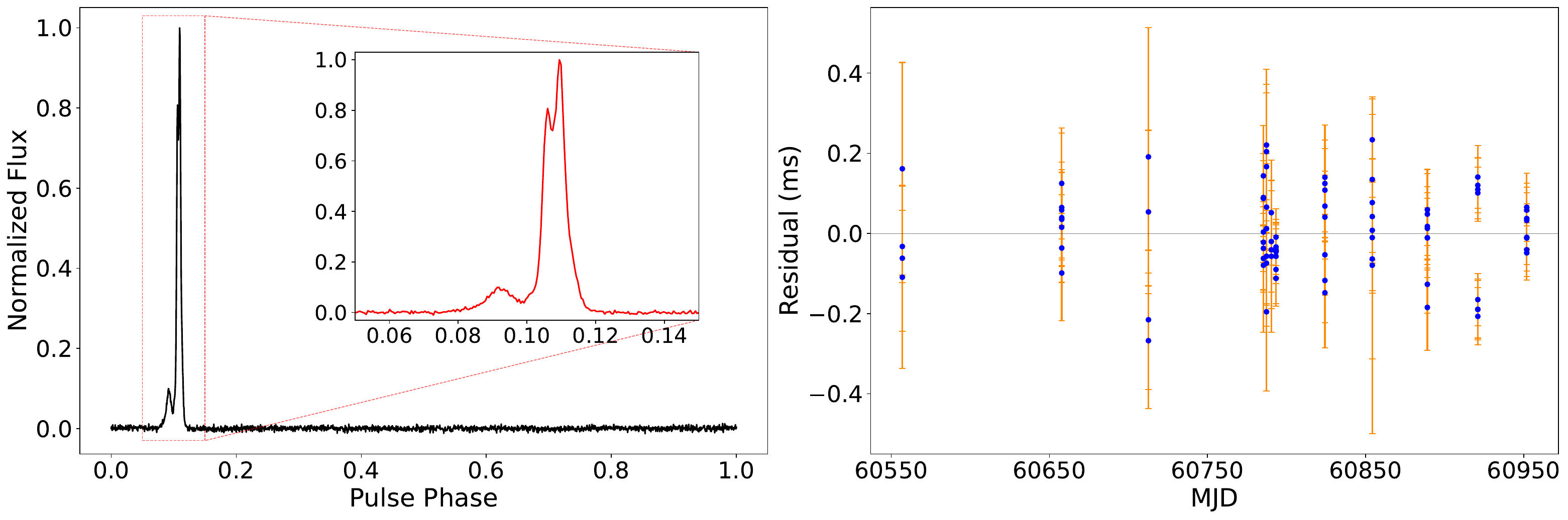}
    \caption{Left panel: The pulse profile with 1.6-hour integration time. Due to the very low duty cycle of the pulse signal, a zoom-in view of the phase range 0.05--0.15 is provided. Right panel: The timing residuals plotted against observational date for the four frequency channels.}
    \label{fig:timingres}
\end{figure}

We present the timing solution for PSR J0630+19 in Table~\ref{tab:timing}, along with the post-fit timing residuals using four frequency sub-bands in the right panel of Fig.~\ref{fig:timingres}. The measured spin period is $P\approx1.25~\rm{s}$, with first period derivative of $\dot{P}\approx1.23\times10^{-17}~\rm{s~s^{-1}}$. These parameters yield a characteristic age of approximately $1.6\times10^{9}$ yr, indicating that this is an old pulsar. The fitted DM, $\sim47.87~\rm{pc~cm^{-3}}$, is slightly smaller than the value obtained from our searching analysis. And the coordinates derived from the timing solution, $\rm{R.A.}=06$:30:05.4770, $\rm{Dec.}=+19$:34:32.99, show an offset of $\sim0.51\arcmin$ from the position we calculated in Section~\ref{sec:data}.

\begin{deluxetable}{cccc}[b]
\tablecaption{The parameters of three beam pointing that detected pulse signals \label{tab:loc}}
\tablehead{
\colhead{Beam Pointing} &
\multicolumn{2}{c}{Coordinate$^{a}$} &
\colhead{SNR$^{b}$} \\
\colhead{} &
\colhead{R.A.} &
\colhead{Dec.} &
\colhead{}
}
\startdata
M01P1 & 6:30:04.00 & 19:37:00.0 & 15.9 \\
M01P4 & 6:29:58.26 & 19:34:30.9 & 24.6 \\
M02P3 & 6:30:11.15 & 19:34:30.3 & 38.1 \\
\enddata
\tablenotetext{a}{The coordinates of beam center of each beam pointing.}
\tablenotetext{b}{The SNR of detected pulses.}
\end{deluxetable}

Fig.~\ref{fig:timingres} shows the 1.6-hour integrated pulse profile of PSR J0630+19, which exhibits a very small duty cycle. Since the pulse profiles cannot be easily  described by Gaussian-like curves, we therefore use the intensity weighted width (IWW) to define the pulse width, i.e. we treat the pulse profile as the temporal intensity distribution function, and calculate the standard deviation of time. A correction factor of $(8\ln 2)^{1/2}$ is applied to the standard deviation when reporting the final pulse width. This factor ensures that the resulting IWW corresponds to the full width at half maximum (FWHM) in the case of a Gaussian profile. Using this method, the IWW of the integrated pulse profile is determined to be 31.41 ms in units of time, with an uncertainty of $\sigma_{\rm{IWW}}=0.03$ ms.

The mean flux density of the radio pulse, denoted as $S_{\rm mean}$, can be calculated using the following expression from \citet{Lynch2011}:
\begin{equation}
S_{\rm mean}=\frac{\alpha \beta T_{sys}}{G\sqrt{N_{\rm p}\Delta \nu T_{\rm int}}}\sqrt{\frac{W}{P-W}}
\end{equation}
where $\alpha$ is the $S/N$ threshold, $\beta$ is the sampling efficiency, $T_{sys}$ is the system temperature, $T_{\rm int}=5848.2$ s is the integration time, and $G$ represents for antenna gain. For FAST, we \textbf{adopted} $\beta\approx1$, $T_{\rm sys}=24$ K, and $G=16$ K Jy$^{-1}$. The bandwidth \textbf{is} $\Delta\nu=336.8$ MHz as we masked $\sim163.2$ MHz in our pipeline due to the RFI. Additionally, $N_{\rm p}=2$ is the number of polarizations, $P$ is the period of pulsar, and $W$ is the width of pulse profile. Here we \textbf{used} IWW, yielding $W/P\approx2.52\%$. The mean flux density at 1.25 GHz \textbf{was} estimated to be 93.0 $\mu$Jy, and the major error of flux comes from the noise temperature variation, which is 20\% as measured in the FAST engineering phase. Approximately $29.2\%$ of pulsars in the ATNF catalogue have a flux density less than 93 $\mu$Jy at 1.4 GHz.

\begin{deluxetable}{lc}[b]
\tablecaption{Timing parameters for PSR J0630+19 \label{tab:timing}}
\tablehead{
\colhead{Parameter} &
\colhead{Value}
}
\startdata
MJD range & 60556--60952 \\
Reference epoch (MJD) & 60712.6 \\
R.A., $\alpha$ (J2000) & 06:30:05.4770(14) \\
Dec., $\delta$ (J2000) & +19:34:32.99(17) \\
Spin frequency, $f$ (Hz) & 0.8009359305558(29) \\
First spin frequency derivative, $\dot{f}$ ($10^{-18}$~Hz~s$^{-1}$) & $-7.91(51)$ \\
DM (pc~cm$^{-3}$) & 47.870(17) \\
Number of ToAs & 84 \\
Residuals rms ($\mu$s) & 95.125 \\
EFAC & 1 \\
\hline
\multicolumn{2}{c}{Derived parameters} \\
\hline
Galactic longitude, $l$~($^\circ$) & 193.1097(1) \\
Galactic latitude, $b$~($^\circ$) & $+4.2890(1)$ \\
Spin period, $P$ (s) & 1.2485393173785(46) \\
First spin period derivative, $\dot{P}$ ($10^{-17}$~s~s$^{-1}$) & 1.234(80) \\
Characteristic age, $\tau_{\mathrm{c}}$ (Myr) & 1604.6(104.0) \\
Surface magnetic field, $B$ ($10^{11}$~G) & 1.26(0.04) \\
Spin-down luminosity, $\dot{E}$ ($10^{29}$~erg~s$^{-1}$) & 2.50(0.16) \\
Distance, $d$ (kpc) & 0.972-1.566\tablenotemark{a} \\
\enddata
\tablenotetext{a}{The distance is derived from the dispersion measure following the Galactic electron density models by \citet{2002astro.ph..7156C} and \citet{2017ApJ...835...29Y}.}
\end{deluxetable}

\section{Discussion}

The “death line” delineates the theoretical boundary beyond which a pulsar's magnetosphere can no longer sustain pair-production cascades. It is defined by the condition that the electrostatic potential drop along a magnetic field line traversing the polar gap is sufficient to generate enough electron-positron pairs to screen the accelerating electric field \citep{1975ApJ...196...51R,1993ApJ...402..264C,2000ApJ...531L.135Z}. Two primary inner gap models account for the binding of ions at the neutron star surface. The traditional vacuum gap model \citep{1975ApJ...196...51R,1992A&A...254..198B}, which forms the basis for the classic death line, assumes that ions are strongly bound to the surface. In contrast, the space-charge-limited flow (SCLF) model \citep{1979ApJ...231..854A,2000ApJ...531L.135Z} assumes the free emission of particles from it. Both models were originally proposed under the postulate that the primary mechanism for generating the high-energy photon 
seeds, which ignite the pair-production cascades, is either curvature radiation (CR) or inverse Compton scattering (ICS, \citet{1996A&A...310..135Z,1997ApJ...478..313Z}). \textbf{Please note that CR cannot contribute to the primary photons responsible
for the TeV emission.} The ``death valley" \citep{1993ApJ...402..264C} is bounded by the theoretical death lines of two pulsar models: one with a pure dipole field and the other with an extreme case of a twisted surface magnetic field, both based on the vacuum gap model.

\begin{figure}[b]
    \centering
    \includegraphics[width=0.9\textwidth]{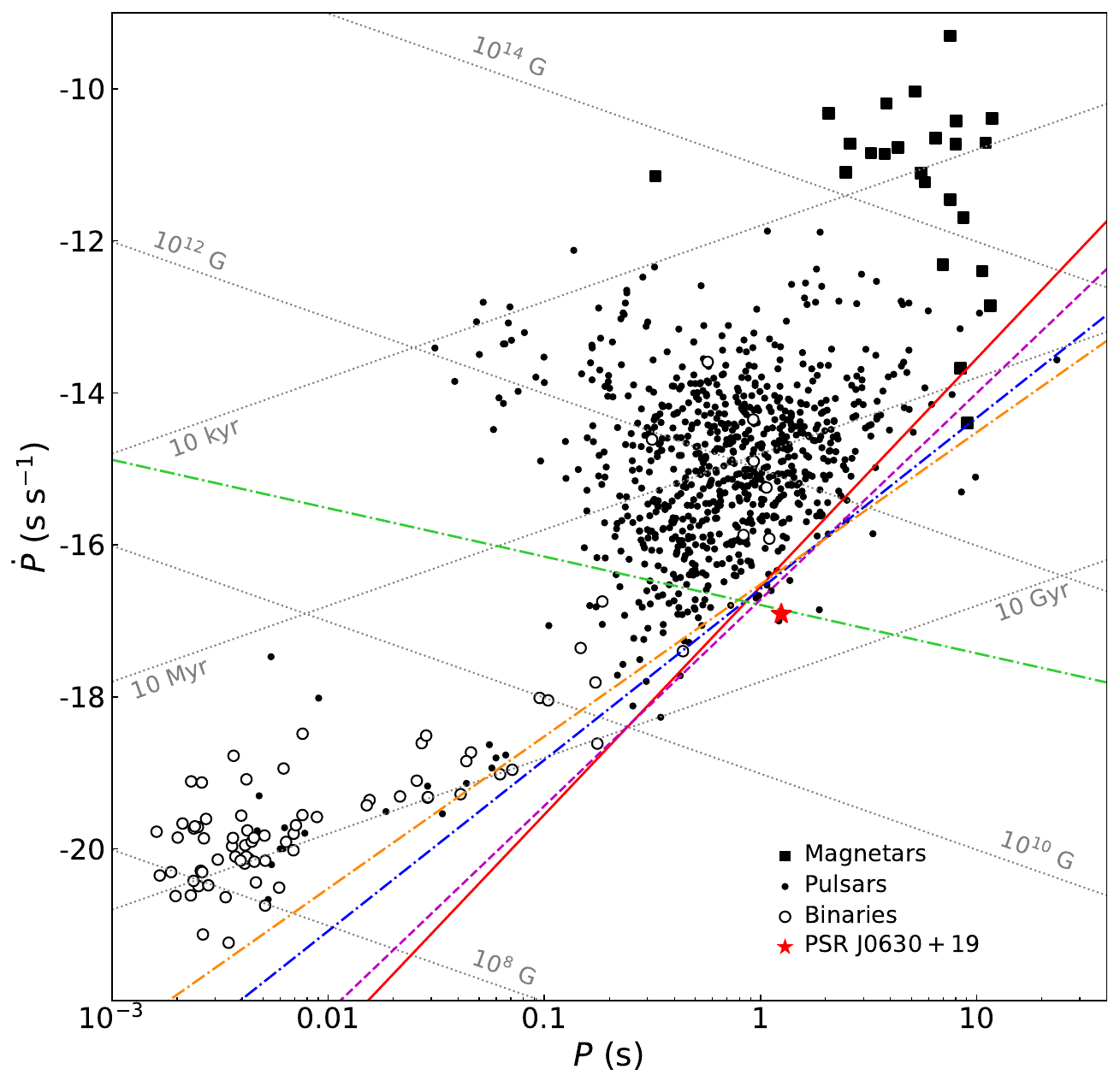}
    \caption{$P$-$\dot{P}$ diagram, with the position of PSR~J0630+19 marked by a red star. Magnetars, normal pulsars, and binary pulsars are represented by black squares, black dots, and black circles, respectively. Several colored lines indicate the theoretical “death lines” below which pulsars are not expected to emit radio signals according to different models. The red solid line corresponds to the traditional death line \citep{1975ApJ...196...51R,1992A&A...254..198B}. The purple dashed line shows the death line from Equation~(9) of \citet{1993ApJ...402..264C}. The blue dash-dotted line is the prediction from curvature radiation in the vacuum gap model \citep{2000ApJ...531L.135Z}. And the orange dash-dotted line represents the death line predicted by curvature radiation (CR) in the space-charge-limited flow (SCLF) model \citep{2000ApJ...531L.135Z}. Finally, the green dash-dotted line corresponds to the death line from inverse Compton scattering (ICS) in SCLF model.}
    \label{fig:ppdot}
\end{figure}

Fig.~\ref{fig:ppdot} presents the $P-\dot{P}$ diagram for the pulsars, with PSR J0630+19 marked by a red star. And several theoretically predicted death lines are also plotted in the figure. The traditional death line of radio pulsars is equivalent to an spin-down luminosity $\dot{E} \approx 10^{30}$ erg s$^{-1}$ \citep{2020RAA....20..188W}, and low $\dot{E}$ cases are usually used to test the pulsar radio emission models. For example, PSR J2144-3933 is a low $\dot{E}$ pulsar with period of 8.5 s and beyond the traditional death line and death valley \citep{1999Natur.400..848Y}. But the \textbf{ICS in the SCLF model, i.e., ICS-SCLF model} proposed by \citet{2000ApJ...531L.135Z} can sustain strong pair production for this long-period pulsar. In this work, PSR J0630+19 is at lower-right of the $P-\dot{P}$ diagram in Fig.~\ref{fig:ppdot} with a long period about 1.2 s. The death line of traditional one (red solid line) defined by \citet{1975ApJ...196...51R} and those investigated by \citet{2000ApJ...531L.135Z} \textbf{including CR in the vacuum gap model (blue dash-dotted line), CR in the SCLF model (orange dash-dotted line) are all above the location of PSR J0630+19. But the pulsar lies on the ICS-SCLF model (green dash-dotted line) derived by \citet{2000ApJ...531L.135Z}. The small discrepancy could be explained by the uncertainties in $P$ and $\dot{P}$.} 

An “observational limit-line” was proposed by \citet{2020RAA....20..188W} based on the lowest flux levels detectable by the FAST telescope, corresponding to a spin-down luminosity of about $\dot{E} \approx 10^{28}$ erg s$^{-1}$. This can \textbf{explain the radio emissions from 
low-$\dot{E}$ pulsars}, such as PSR J2144–3933 \citep{1999Natur.400..848Y,2020RAA....20..188W} and 
PSR J0211+4235 \citep{2023MNRAS.522.5152W}. An upper limit in the efficiency ($\xi_{\rm max} = 0.01$) of radio emission was proposed to correspond to the death \textbf{line of 
low-$\dot{E}$ pulsars} \citep{2014ApJ...784...59S}. The radio emission efficiency is defined as the fraction of the rate of loss of rotational energy transformed into radio emission ($\xi\equiv \frac{L}{\dot{E}}$, where $L$ is the radio luminosity from the integrated radio energy flux). The spin-down power $\dot{E}$ is derived from pulsar timing measurements. The radio luminosity $L$ is derived \textbf{from measuring the radio} emission flux and the pulsar distance. The radio efficiency is different from X-ray (e.g. \citet{2013arXiv1304.4203S}) or $\gamma$-ray (e.g. \citet{2013ApJS..208...17A}) ones. A near-linear inverse correlation between $\xi$ and
 the spin-down power, $\xi \propto \dot{E}^{-0.9}$ for normal pulsars was \textbf{reported} by \citet{2014ApJ...784...59S}.
 In this work, the radio luminosity of PSR J0630+19 is $L\approx 6.5 \times 10^{26}$ erg s$^{-1}$ based on the Eq (2) in the work of \citet{2014ApJ...784...59S} with FAST measured mean flux density at 1.25 GHz and the pulsar distance \citep{2017ApJ...835...29Y} to the Earth. Combining with the $\dot{E}$ \textbf{given} in Table~\ref{tab:timing}, we can derive the radio efficiency for this pulsar \textbf{as} $\xi \approx 2.6\times 10^{-3}$, much lower than the upper limit of $\xi$ for low-$\dot{E}$ pulsars. 

The GeV $\gamma$-ray conversion efficiency is defined by $\eta\equiv L_{\gamma}/\dot{E}$, where $L_{\gamma}$ is the gamma-ray luminosity derived from the observed GeV $\gamma$-ray flux and distance $d$ to the pulsar, i.e., $L_{\gamma} = 4 \pi d^2 f_{\Omega}S_\gamma$, where $S_\gamma$ is integrated energy flux in observed GeV band ranges. And $f_{\Omega}$ is the beam correction factor \citep{2010ApJ...714..810R}  to extrapolate the observed flux to the full sky for beam shape model. In the magnetic and rotational model of the neutron star, the factor is a function of the inclination and viewing angles. 
\citet{2023ApJ...954..204K} 
\textbf{found} $f_{\Omega} < 1$ for the LAT pulsar sample. 
This value corresponds to the specific scenario of a fan-like beam from the outer magnetosphere that sweeps the entire sky. A linear correlation between $L_{\gamma}$ and the open field-line voltage gives $L_{\gamma} \propto \sqrt{\dot{E}}$ \citep{1996A&AS..120C..49A}. The observed $\gamma$-ray luminosity $L_{\gamma}$ in the 0.1–100 GeV energy band by the Large Area Telescope (LAT) on the Fermi Gamma-ray Space Telescope was found to be highly
correlated with the available pulsar power \citep{2023ApJ...958..191S}. A tight correlation between the pulsar GeV emission and the associated PWN TeV
emission was observed \citep{2015ApJ...804...25A}, which suggests the possibility of a linear relationship between the two emission
mechanisms. A systematic study of the population of TeV PWNe observed by \textbf{H.E.S.S. found} a power-law relation between TeV
luminosity $L_{TeV}$ and pulsar spin-down power $\dot{E}$ as $L_{TeV} \sim \dot{E}^{0.58\pm0.21}$, \textbf{consistent with 
the model \citep{2012arXiv1202.1455M} that suggests
a power index of around 0.5 \citep{2018A&A...612A...2H}}. 
The work of \citet{2020ApJ...905...76A} reported the best-fit fluxes and spectral indices of the 3HWC sources. These fits assume a power-law spectrum,
\begin{equation}
dN/dE = F_0(E/E_0)^\alpha ,
\end{equation}
where $E_0$ is a reference energy, $F_0$ is the differential flux at $E_0$,
and $\alpha$ is the spectral index. The reported spectral indices of the 3HWC sources are
interpreted as an average or effective spectral index across HAWC’s energy range [$E_1, E_2$] TeV. The integrated TeV energy flux 
$S_{TeV}$ in the TeV 
energy band measured at Earth is obtained by
\begin{equation}
S_{TeV} = \int^{E_2}_{E_1} E \frac{dN}{dE} dE.
\end{equation}
\textbf{the parameters of best-fit spectrum and energy range for 3HWC J0630+186 were presented in Table 2 of the work of \citet{2020ApJ...905...76A}. We} calculated the integrated TeV energy flux $S_{TeV} \approx 8.74^{+3.87}_{-2.27} \times 10^{-13}$ erg s$^{-1}$ cm$^{-2}$ based on Eq. (3). If the distance of 3HWC J0630+186 is the same as that of PSR J0630+19 showed in Table~\ref{tab:timing} and the TeV efficiency, conventionally defined as the ratio of $L_{TeV}$ to $\dot{E}$, is assumed to be unity, we can obtained its TeV $\gamma$-ray luminosity \textbf{as} $L_{TeV}\approx 1.7^{+1.1}_{-0.9} \times 10^{32}$ erg s$^{-1}$, 
which is much larger than the spin-down luminosity $\dot{E}$ of PSR J0630+19. This implies that the TeV $\gamma$-ray emissions of 3HWC J0630+186 is not powered by spin-down power of PSR J0630+19. The apparent conversion of spin-down power into $\gamma$-ray flux in excess of 100\% can be attributed to an underestimated distance $d$ and/or an erroneous assumed TeV efficiency. 
Future distance measurement and pulsar searching around 3HWC J0630+186 
may help to understand the origin of its very high energy emissions. 



\section{Summary} \label{sec:conc}

PSR J0630+19 had been considered the only plausible counterpart to the TeV source 3HWC J0630+186. We observed \textbf{this pulsar utilized FAST from September 4, 2024 to October 4, 2025, to find out if both the TeV source and the pulsar} are related or not. We \textbf{detected the pulse period of the pulsar as} the same as that in previous observations reported by Arecibo telescope. Using the FAST observations conducted in the $SnapShotDec ~Mode$, we \textbf{derived} a new, more precise localization for PSR J0630+19.
Timing solution obtained by FAST tracking Mode for PSR J0630+19 \textbf{showed} that it is an old pulsar with \textit{spin-down energy luminosity $(2.50 \pm 0.16) \times 10^{29}$ erg s$^{-1}$.}  The spin-down power and radio luminosity of the pulsar yield a radio efficiency that places it within the region predicted for the lowest flux levels detectable by the FAST telescope. Adopting the distance to PSR J0630+19 derived from its DM and the HAWC spectral parameters, we calculated the TeV $\gamma$-ray luminosity of 3HWC J0630+186. This $\gamma$-ray luminosity significantly exceeds the spin-down power of PSR J0630+19, consequently ruling it out as the power source for 3HWC J0630+186. Therefore, the very-high-energy emission most likely originates from another, yet-unidentified source or physical process. Future multi-wavelength observations targeting the region of 3HWC J0630+186 will help \textbf{to understand} its true origin.

\begin{acknowledgments}
This work made use of the data from FAST (Five-hundred-meter Aperture Spherical radio Telescope; https://cstr.cn/31116.02.FAST). FAST is a Chinese national mega-science facility, operated by National Astronomical Observatories, Chinese Academy of Sciences. We would like to thank the FAST group and PKU pulsar group for the suggestions about data analysis and helpful suggestions that led to significant improvement in our study. This work is supported by National SKA Program of China (2025SKA0140100), China Postdoctoral Research Foundation No.~2023M743516, National Natural Science Foundation of China (12403093). Project Supported by the Specialized Research Fund for State Key Laboratory of Radio Astronomy and Technology.
\end{acknowledgments}

\begin{contribution}
X.-H.C. and B.-J.W. proposed the observational project. The FAST team 
designed and
scheduled the observations during the FAST commissioning stage. 
B.-J.W. analysed the data. 
B.-J.W. and X.-H.C. wrote the paper. H.X., J.-G.L. and R.-X.X. participated in the interpretation of the results.
All authors discussed the contents of the paper and contributed to the preparation of the manuscript.


\end{contribution}

\bibliography{sample701}{}
\bibliographystyle{aasjournalv7}



\end{document}